\documentclass[seceq,supplement]{ptptex}

\usepackage{graphicx}
\usepackage{wrapft}

\title{
Study of Transient Nuclei near Freezing
}

\author{Masaharu \textsc{ISOBE$^1$} and Berni J. \textsc{ALDER$^2$}
}

\inst{$^1$Graduate School of Engineering, Nagoya Institute of Technology, Nagoya, 466-8555, Japan; 
$^2$Lawrence Livermore National Laboratory, P.O.Box 808, Livermore, CA 94551-9900, USA
}

\abst{
The molasses tail in dense hard core fluids is investigated by extensive event-driven molecular dynamics simulation through the orientational autocorrelation functions.
Near the fluid-solid phase transition, there exist three regimes in the relaxation of the pair orientational autocorrelation function, namely the kinetic, molasses (stretched exponential), and diffusional power decay.
The  density dependence of both the molasses and diffusional power regimes are evaluated and the latter compares with theoretical predictions in three dimensions.
The largest cluster at the freezing density of only a few sphere diameter in size persist for only about 30 picoseconds ($\sim 2.8 \times 10^{-11}$[s]).
The most striking observation through the bond orientatinal order parameter is the dramatic increase of the cluster size as the freezing density is approached.
}

\begin{document}

\maketitle

\section{Introduction}

The long slow decaying potential part of the shear-stress autocorrelation function (SACF) has been called the ``molasses tail'' to differentiate it from the hydrodynamic origin of the long time tail in the velocity autocorrelation function~\cite{alder_1967,isobe_2008} and to emphasize its relation to the highly viscous glassy state~\cite{alder_1986,dorfman_1986}.
The decay of SACFs have been well investigated by mode coupling theory (MCT) and kinetic theory, both leading to the same power form as in the case of the velocity autocorrelation function. This applies to the kinetic part of the shear autocorrelation function, but not to the potential part which is central to the molasses tail.
From the numerical point of view, the long time tail of SACF has been computed by several molecular dynamics (MD) simulations in the early 80s; Evans (1980)~\cite{evans_1980}, Wood \& Erpenbeck(1981)~\cite{wood_1981},and Morriss \& Evans(1985)~\cite{morriss_1985}.
These studies seem to show that the long time tail of both the kinetic and potential parts have a power decay consistent with MCT, however, the amplitude of SACF in dense fluids was found to be orders of magnitude greater than predicted.
This discrepancy was ascribed to the possibility that the numerical results were not run long enough. However, we show that the tail is caused by slow structural relaxation in the dense liquid around the peak of the structure factor rather than by hydrodynamic phenomena at long wave length.
This was the starting point of the theory of glassy transition~\cite{goetze_1992,goetze_2008}.
Since then many papers focusing on density fluctuations in the study of the glass transition have been published, however, the microscopic mechanism of the stress field relaxation have not been clarified yet.

Twenty years ago, Ladd and Alder have speculated that the long time tail of the shear stress auto-correlation function near the solid-fluid transition point in the hard sphere system is due to transient crystal nuclei formation~\cite{ladd_1989}.
They found that the potential part of the SACF and the angular orientational auto-correlation function (OACF) are identical in the long time limit and show non-algebraic decay in time.
Since the evidence suggested that the reason for non-algebraic decay is structural relaxation rather than hydrodynamic flow, an attempt was made to understand this slow decaying mechanism by decomposing the OACFs into two-, three-, and four-body correlations, however, many of these correlation functions, especially the four-body correlations, have not been obtained accurately due to the computer limitation at that time.
Even with today's much better computers, it is not possible to get accurate information on the four body distribution and hence we made an attempt to use what could be learned about the cluster size from the bond orientational parameter.
The results, presented here, look promising, and we plan to extend, this nearest neighbour bond order parameter to further neighbours.
The prediction of the cooling rate necessary to prevent crystallization requires knowledge of the rate of growth of a cluster the size of a critical solid nucleus and the time it exists in this transient state.
Therefore, analysis of this slow decaying process of the OACF and its decomposition into several components is expected to be a key factor in understanding the onset of the glass transition~\cite{alder_1986, dorfman_1986}.

Previously~\cite{isobe_2009}, we have reported on a two dimensional system consisting of elastic hard disks at a single density near the solid-fluid transition point placed in a square box with periodic boundary conditions, using a modern fast algorithm based on event-driven MD simulation~\cite{isobe_1999}.
We confirmed the non-algebraic decay (stretched exponential) at intermediate times presumably due to the existence of various sized solid clusters at high densities decaying at different rates.
We also determined the length of time for which the biggest such nuclei exists.
In this study, we focus on the rapidly increasing time with increasing density for the decay of OACFs and are able to establish the length of time for which the biggest such nuclei exists at each density.
We also compare the results to theoretical predictions in the subsequent power law decay.
We will further report results for hard spheres for different particle numbers.
We use a more efficient program code for calculating pair contributions to the OACFs, allowing for more accurate results.

\section{Decomposition of the Orientational Autocorrelation Function}

We focus on the potential part of the SACF $\langle J^P_{xy}(t)J^P_{xy}(0) \rangle$, where $J^P_{xy}$ is the potential part of the momentum current $J^P_{xy}$, where the molasses tail appears.
Since velocities and positions are no longer correlated beyond a few mean collision times, only the orientational part of SACF, namely the OACF $\langle O_{xy}(t)O_{xy}(0) \rangle$ needs to be studied~\cite{ladd_1989, isobe_2009}.
$O_{xy}(t)$ is defined as 

\begin{equation}
O_{xy}(t)=\sum_c \frac{x_{ij}y_{ij}}{\sigma^2}\delta{(t-t_c)},
\label{eqn:1}
\end{equation}

\noindent
where $\sum_c$ means contribution at the colliding time $t_c$ at which $(x_{ij},y_{ij})=(x_i-x_j,y_i-y_j)$ are the relative positions between hard spheres (disks) $i$ and $j$.

To avoid the delta function singularity of $O_{xy}(t)$ for hard particles, the alternative Einstein-Helfand expression~\cite{helfand_1960,alder_1970} involving the second derivative, obtained by the numerical differentiation with five point stencils, is adopted for calculating the correlation function,

\begin{equation}
C(t)=\langle O_{xy}(t)O_{xy}(0) \rangle = \frac{1}{2}\frac{d^2}{dt^2}\langle (G(t)-G(0))^2 \rangle,
\label{eqn:2}
\end{equation}

\noindent
where

\begin{equation}
G(t)=\sum_c \frac{x_{ij}y_{ij}}{\sigma^2} \Theta (t-t_c),
\label{eqn:3}
\end{equation}
\noindent
and where $\Theta (t)$ is the unit step function.
There are three independent orientational factors ($O_{xy}$, $O_{yz}$, and $O_{zx}$) in 3D.
We introduce the reduced (e.g. Enskog values) orientational function $C^*(s)$ in terms of the reduced time $s = t/t_0$,

\begin{equation}
C^*(s) = \frac{\pi m \sigma^2 t_0}{V\eta_E}C(t),
\label{eqn:4}
\end{equation}

\noindent
where $t_0$ is the mean free time and $\eta_E$ is the Enskog shear viscosity.
For a hard core fluid, the Enskog shear viscosity can numerically be estimated by the expression of Gass(1971)~\cite{gass_1971} with the third Sonine polynomial approximation in 2D and Wainwright(1964)~\cite{wainwright_1964} with the equation of state by Ree\&Hoover(1967)~\cite{ree_1967} in 3D.

The total correlation function $C(t)$ can be decomposed into pair  $C_2(t)$($ij-ij$ collision), triplet $C_3(t)$ ($ij-ik$ collision), and quadruplet $C_4(t)$ ($ij-kl$ collision) contributions~\cite{ladd_1979,isobe_2009}, where $i,j,k,l$ are particle index.

\begin{equation}
C(t)=C_2(t)+C_3(t)+C_4(t).
\end{equation}

\noindent
The pair contribution $C_2(t)$ is defined as,

\begin{equation}
C_2 (t) = \frac{1}{N} \left\langle \sum_i^N \sum_{j(j>i)}^N O_{xy}^{ij}(t) O_{xy}^{ij}(0) \right\rangle =  \frac{1}{2N}\frac{d^2}{dt^2} \left\langle \sum_i^N \sum_{j(j>i)}^N G^{ij}(t)^2 \right\rangle,
\label{eqn:5}
\end{equation}

\noindent
since $G^{ij}(0)=0$ ($\dot{G^{ij}}(t)=O^{ij}_{xy}(t)$).
To calculate $C_2$, we introduce a ``collision pair index'' $c_k=(c_i-1)N-c_i(c_i-1)/2+c_j-c_i$, where $N$ is the number of particles, which identifies a given pair quickly, avoiding having to check whether the collision pair is same as before in the process of calculating $G^{ij}(t)$.
This speeds up the calculation considerably.
\footnote{For example, in case of $N=4$, the total number of 
collision pairs are $N(N-1)/2=6$, which can be listed as $(c_i,c_j)=(1,2),(1,3),(1,4),(2,3),(2,4),(3,4)$, where $c_i < c_j$.
By using the ``collision pair index'', we obtain $c_k=1,2,3,4,5,6$ for each collision pair, respectively.
Therefore, it is easy and convenient to deal with the collision
pair as the sequential number to sort and insert the array of 
correlation pair $G^{ij}(t)$ when we make the computer program.}

\section{Results}

The system consists of hard disks or hard spheres placed in a $L_x \times L_y$($=A$) square box or a $L_x \times L_y \times L_z$($=V$) cubic box, respectively, with periodic boundary conditions.
Initially, the simulation systems for each packing fraction $\nu$ $(= N \pi (\sigma/2)^2/A$ for 2D, or $(4/3) N \pi (\sigma/2)^3/V$ for 3D), are prepared as the equilibrium state by a sufficiently long preliminary run.
Relative to the close packed area $A_0$, and volume $V_0$, $\nu=\pi/(2\sqrt{3}(A/A_0))$ and $\nu=\pi/(3\sqrt{2}(V/V_0))$, respectively.
The system evolves through collisions, using an algorithm based on event-driven Molecular Dynamics (MD) simulation~\cite{isobe_1999}.
Most of the calculations are done with the particle number $N=512$ and $4,096$.
The density is varied at relatively dense value near the solid-fluid transition point, which are known to be $\nu_c \sim 0.70$ for 2D and $\sim 0.48$ for 3D.
Ladd \& Alder (1989)~\cite{ladd_1989} have calculated OACFs and its separate parts at several densities; $A/A_0 = 1.35, 1.4, 1.6$ for 2D and $V/V_0 = 1.5, 1.6, 1.8, 2.0$ for 3D.
These densities correspond to $\nu = 0.672\cdots, 0.648\cdots, 0.567\cdots$ for 2D and $\nu = 0.493\cdots, 0.463\cdots, 0.411\cdots, 0.37\cdots$ for 3D, respectively.
In our calculation, we use several additional packing fractions up to solid-fluid transition density; $\nu = 0.69, 0.67, 0.65, 0.57$ for 2D and $\nu = 0.47, 0.45, 0.40, 0.35$ for 3D.

\subsection{Density \& system size dependence of OACFs in 2D and 3D}

First, we calculate the {\em total OACFs} for different packing fractions and system size to investigate the functional form of the molasses tail at long times, and the effects of periodic boundary conditions.
In Figs.~1 and 2, the dimensionless OACFs for various packing fractions near the solid-fluid transition point in terms of reduced time $s=t/t_0$ for 2D and 3D are shown, respectively.
Statistical averages are made over $10^{11}$ to $10^{12}$ total collisions of event-driven MD runs, which are $10^3 \sim 10^4$ larger than previous work~\cite{ladd_1989}.
Typical error bars are also shown at long times.
Two different number of particles are shown in both 2D and 3D to observe periodic boundary effect due to sound wave propagation across the system.
The sound wave propagation time explicitly appears in the long time tail of velocity auto-correlation functions (e.g., Ref.~\cite{isobe_2008}), but appears to have no effects here, as expected, since OACF does not involve velocities.

\begin{figure}
\center
\includegraphics[scale=0.50]{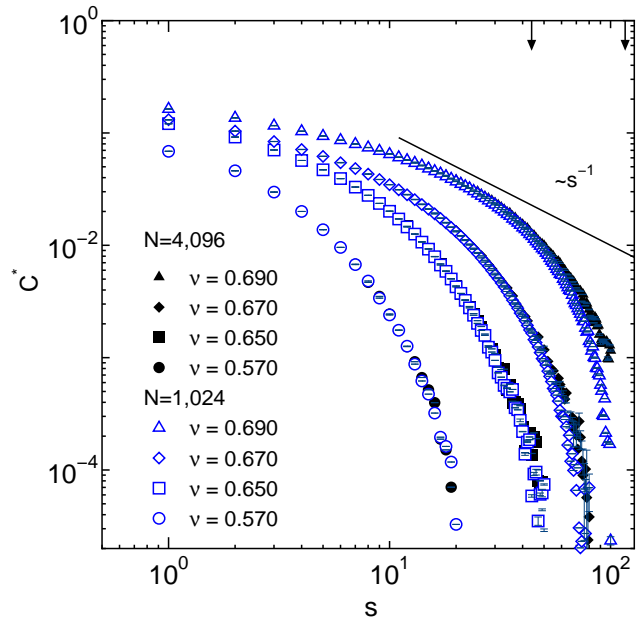}
\caption{
A log-log plot of the dimensionless OACFs (see Eq.(2.4)) for various packing fractions for two different sized systems in 2D versus reduced time, $s=t/t_0$ are shown. Arrows describe the approximate values of the sound wave transversal time (the left for $N=1024$ and the right for $N=4096$.)}
\end{figure}

\begin{figure}
\center
\includegraphics[scale=0.50]{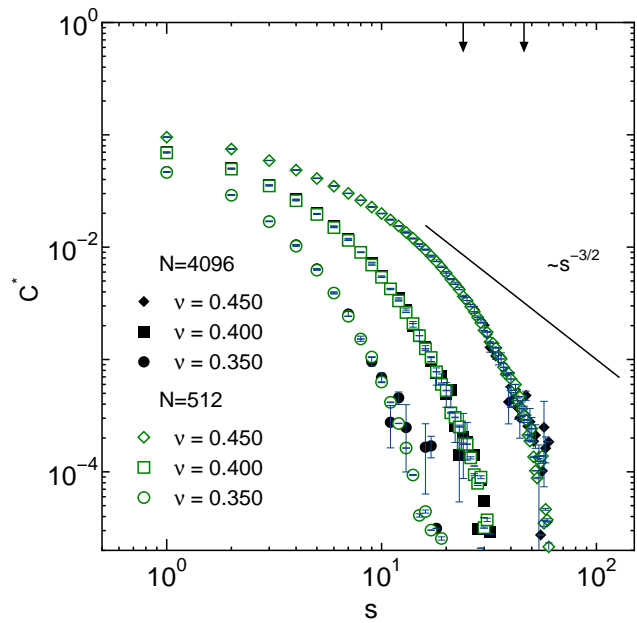}
\caption{
As Fig.1 except for 3D.}
\end{figure} 

\noindent
The OACFs for both 2D and 3D clearly show the slower decay when the system become denser and their decay are not the power law form as for the hydrodynamic tail $\sim s^{-d/2}$, where $d$ is dimension.
These data suggest that the theoretical prediction of MCT for the long time tail must be reconsidered in dense liquids.
Based on these results, we choose the relatively smaller system size $N=4096$ in 2D and $N=512$ in 3D for efficiency to calculate $C_2$ in the following subsection.

\subsection{Density dependence of $C_2$ in 2D and 3D}

To investigate the density dependence of $C_2$ up to the time of the diffusional power regime, long runs were performed for various packing fractions.
Since the maximum collision index (i.e., $c_k=N(N-1)/2$) increases as $\sim {\cal O}(N^2)$, it is difficult to calculate correlation functions averaged for all possible collision pairs $c_k$ due to the restriction on CPU time and memory, if $N$ becomes more than one thousand particles.
For instance, one needs about $100Mbytes$ of memory for the calculation of $C_2$ for all possible collision pairs in the $500$ particles system, comparable to the Ladd \& Alder(1989) calculation, which presented no serious problem.
However, in the case of $N=4,096$ the maximum collision index becomes $c_k= 8,386,560$, which needs about $10Gbytes$ of main memory.
To reduce the calculation to a manageable one, we restrict the range of collision pairs $c_k$ from $1$ to $10 N$ in the averaging procedure.
This restriction of $c_k$ only affects to error bars, but we perform sufficiently long time MD runs for $C_2$, much longer than the typical time scale of the largest cluster breaking up.

\begin{figure}
\begin{minipage}{0.47\hsize}
\begin{center}
\includegraphics[scale=0.42]{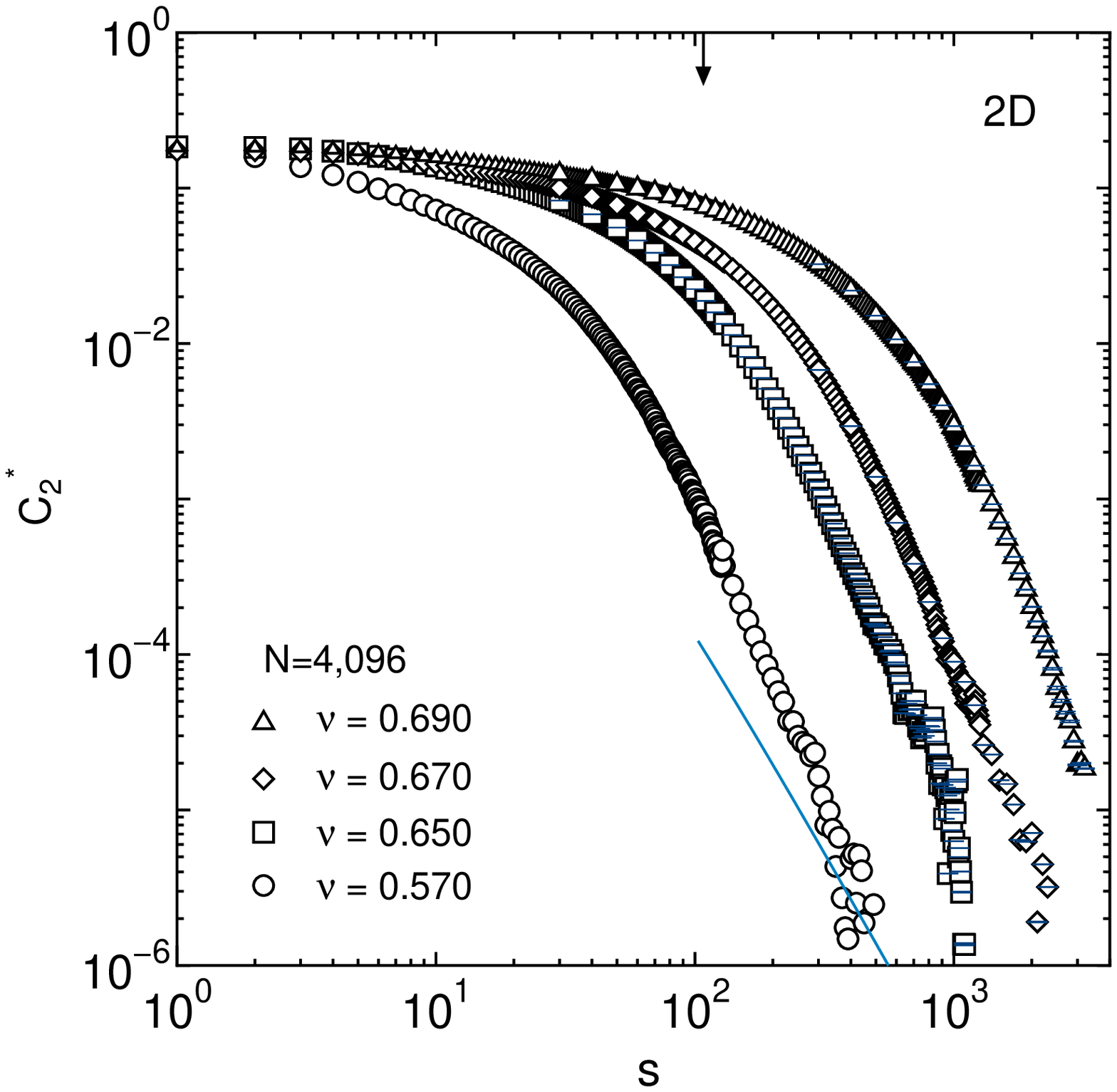}
\end{center}
\end{minipage}
\hspace{2mm}
\begin{minipage}{0.47\hsize}
\begin{center}
\includegraphics[scale=0.42]{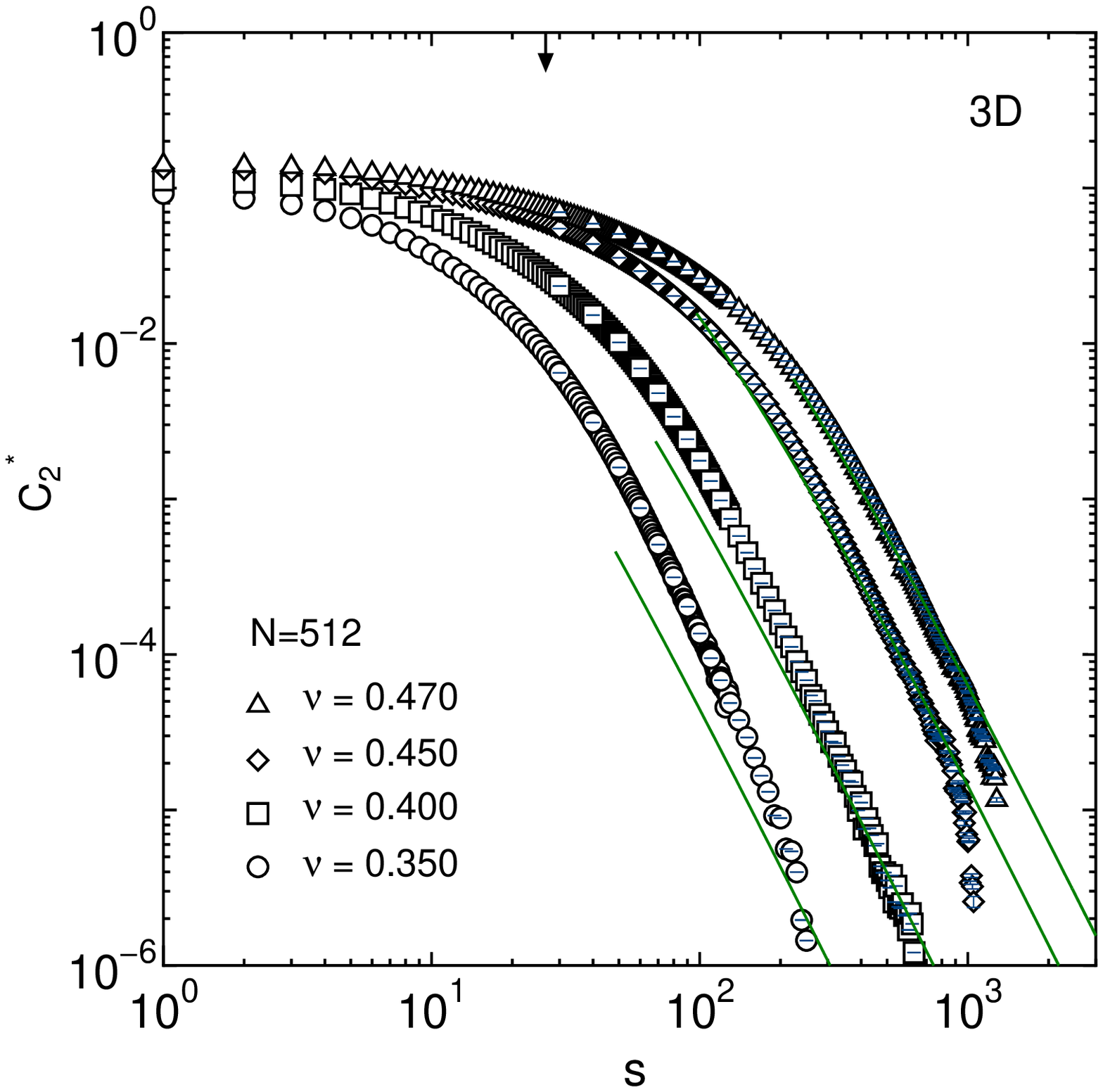}
\end{center}
\end{minipage}
\caption{
$C^*_2$s for various packing fractions near the solid-fluid transition point in 2D (left) and 3D (right) in terms of reduced time $s=t/t_0$.
Arrows indicate the sound wave transversal time for each system size.
Theoretical slopes by using the effective diffusion constant at $\nu=0.57$ in 2D(left) and the diffusion constant obtained from actual simulation for each packing fraction in 3D(right) at long times are given by the lines (see, Sec.~3.4).
}
\end{figure}

\noindent
We performed up to $10^{12}$ collision for several independent runs for different time resolution, namely $\Delta t = t_0, 10t_0$, and $100t_0$ at each packing fraction.
We confirmed that these runs give consistent results.
The efficiency of these calculation in such that it does not significantly slow down the MD calculation.
The error bars in the figures are smaller than the symbols except at a few points at long times.
As indicated in Fig.~3, $C_2$ in both 2D and 3D decays slower as the density becomes higher.
The results are consistent with both our previous method at $\nu=0.65$~\cite{isobe_2009} and Ladd \& Alder(1989)~\cite{ladd_1989}.
It is interesting to observe that qualitatively the decay is similar for 2D and 3D.
Next, we investigate the decay of $C_2$ in the two separate time regime, namely the molasses and diffusional regimes.

\subsection{Molasses regime in $C_2$}

\begin{figure}
\center
\includegraphics[scale=0.45]{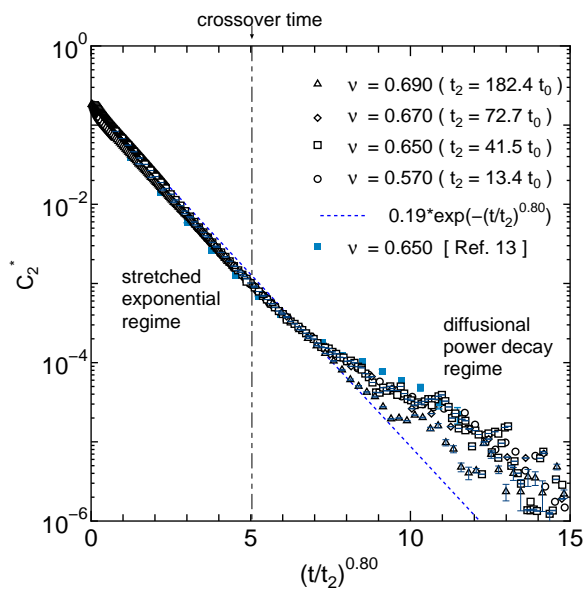}
\caption{Decay of the $C^*_2$ in a semilog plot for $4096$ particles against the scaled time $(t/t_2)^{0.8}$ for the several packing fractions in 2D.}
\end{figure}

\begin{figure}
\center
\includegraphics[scale=0.45]{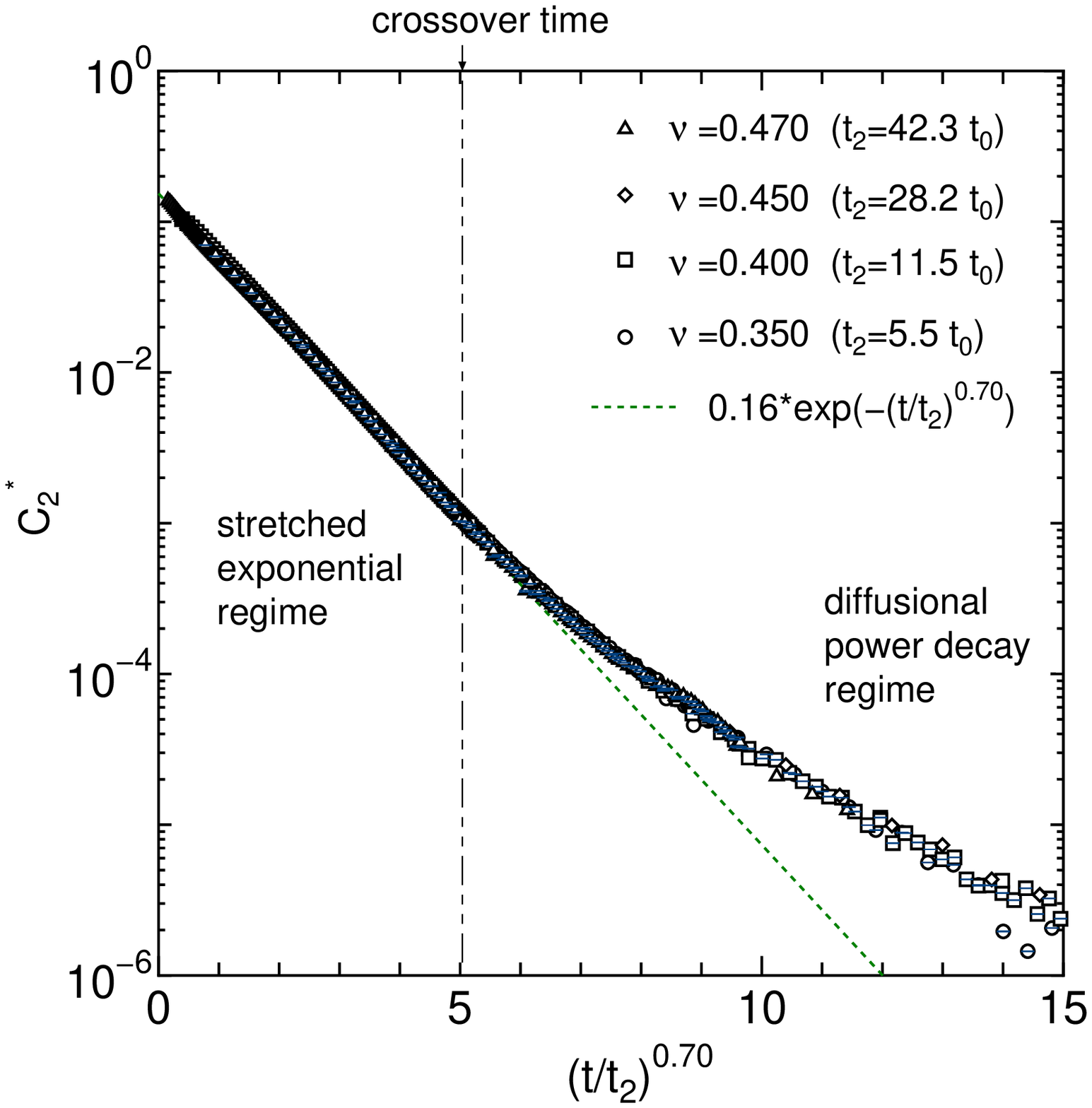}
\caption{Decay of the $C^*_2$ in a semilog plot for $512$ particles against the scaled time $(t/t_2)^{0.7}$ for the several packing fractions in 3D.}
\end{figure} 

We investigate the scaling behavior in the molasses regime as we did previously~\cite{ladd_1989}.
In the molasses regime, $C^*_2$ was found to have a scaling law of the stretched exponential form as,

\begin{equation}
C^*_2(t)=B\exp{\left[-\left(\frac{t}{t_2}\right)^\alpha\right]},
\label{eqn:16}
\end{equation}

\noindent
where $t_2$ is the relaxing time in the molasses regime and $\alpha$ is a density ``independent'' stretched exponent.
The value of the exponent $\alpha$ was found to be $\sim 0.80$ for 2D and $\sim 0.70$ for 3D in previous work~\cite{ladd_1989}.
By adopting a simple fitting method to $C_2$ as given by eq.~(\ref{eqn:16}) with the given values of $\alpha$, we estimated $t_2$.
$C_2^*$ for the several packing fractions plotted against the scaled time $(t/t_2)^\alpha$ are shown in Figs.~4 and 5.
We found that the curves for various packing fraction collapse to a straight line in the regime of $(t/t_2)^\alpha$ between $1$ and $5$.
Thus, the decay of $C_2^*$ is in good agreement with the stretched exponential form of eq.~(\ref{eqn:16}), which is a similar result given in Ref.~\cite{ladd_1989}.
The crossover time $t_{cs}$ from the molasses regime to the diffusional power regime in terms of the mean free time $t_0$ are summarized in Table I and Fig.~6.

\begin{table}
\begin{center}
\begin{tabular}{cccccc} \hline \hline
packing fraction(2D) & $\nu$  & 0.57 & 0.65 & 0.67 & 0.69 \\
crossover time & $t_{cs}/t_0$ & 100 & 287 & 536 & 1364 \\ \hline
packing fraction(3D) & $\nu$  & 0.35 & 0.40 & 0.45 & 0.47 \\
crossover time & $t_{cs}/t_0$ & 55 & 115 & 281 & 431 \\ \hline \hline
\end{tabular}
\end{center}
\caption{
The crossover time $t_{cs}$ from stretched exponential to power form in terms of the mean free time $t_0$ for various packing fractions are shown.}
\label{table:1}
\end{table}

\begin{figure}[ht]
\center
\includegraphics[scale=0.40]{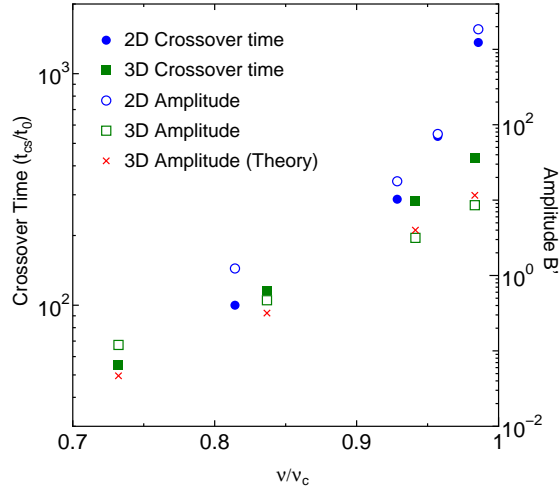}
\caption{
The log of the crossover time $t_{cs}$ divided by the mean free time $t_0$ and the log of the amplitude $B'$, both as a function of the packing fraction given relative to the freezing point. 
}
\label{fig:cross}
\end{figure}

\noindent
The log of the crossover time is increasing linearly in 3D, but faster than exponential in 2D with packing fraction near the solid-fluid transition point $\nu_c$.

\subsection{Diffusional regime in $C_2$}

In the regime of $(t/t_2)^\alpha > 6$, the data show a different functional form from the molasses regime, in which the correlation function changes from a stretched exponential to a power decay, 

\begin{equation}
C^*_2(t)=B' t^{-\beta},
\end{equation}

\noindent
where $\beta$ and $B'$ are the exponent and the amplitude, respectively.
From past theoretical work based on MCT and the pair diffusion mechanism, it is predicted that $\beta=7/2$ in 3D.
Leegwater\&Beijeren(1989)~\cite{leegwater_1989} investigated the equation for two particles diffusing relative to each other, which predicts at long times the behavior of $C_2$ as

\begin{eqnarray}
C^*_2(t) & = & \frac{2\pi n t_E}{15\beta \eta_E} (n \sigma^3 g(\sigma)) \left(\frac{\sigma^2}{4Dt}\right)\exp{\left(-\frac{\sigma^2}{4Dt}\right)}I_{5/2}\left(\frac{\sigma^2}{4Dt}\right) \\
 & \simeq & \frac{2 \sqrt{2\pi} n^2 t_E \sigma^3 g(\sigma)}{225 \beta \eta_E} \left(\frac{\sigma^2}{4D}\right)^\frac{7}{2} t^{-\frac{7}{2}} \quad \quad (t \rightarrow \infty),
\end{eqnarray}

\noindent
where $n=N/V$ is the number density, $D$ is diffusion constant, $E$ stand for Enskog, $g(\sigma)$ is the pair distribution function at contact, and $I_{5/2}$ is a modified Bessel function. 
They also found that a modified theoretical prediction by incorporating recollision contribution and a potential of mean force is in excellent agreement with the MD data throughout the intermediate-time and long-time regimes $t/t_0 >10$, although this requires an effective diffusion constant about 10\% less than for a single particle~\cite{leegwater_1989}.
By the same theoretical argument applied to the 2D~\cite{wakou_2010}, we obtain 

\begin{eqnarray}
C^*_2(t) & = & \frac{\pi nt_E}{8\beta \eta_E} (n \sigma^4 g(\sigma)) \left(\frac{\sigma^2}{4Dt}\right)\exp{\left(-\frac{\sigma^2}{4Dt}\right)}I_{2}\left(\frac{\sigma^2}{4Dt}\right) \\
 & \simeq & \frac{\pi n^2 t_E \sigma^4 g(\sigma)}{64 \beta \eta_E} \left(\frac{\sigma^2}{4D}\right)^3 t^{-3} \quad \quad (t \rightarrow \infty),
\end{eqnarray}

\noindent
In 2D, the theoretical decay is $C_2(t) \sim t^{-3}$ in the long time limit.

\begin{figure}
\begin{minipage}{0.47\hsize}
\begin{center}
\includegraphics[scale=0.42]{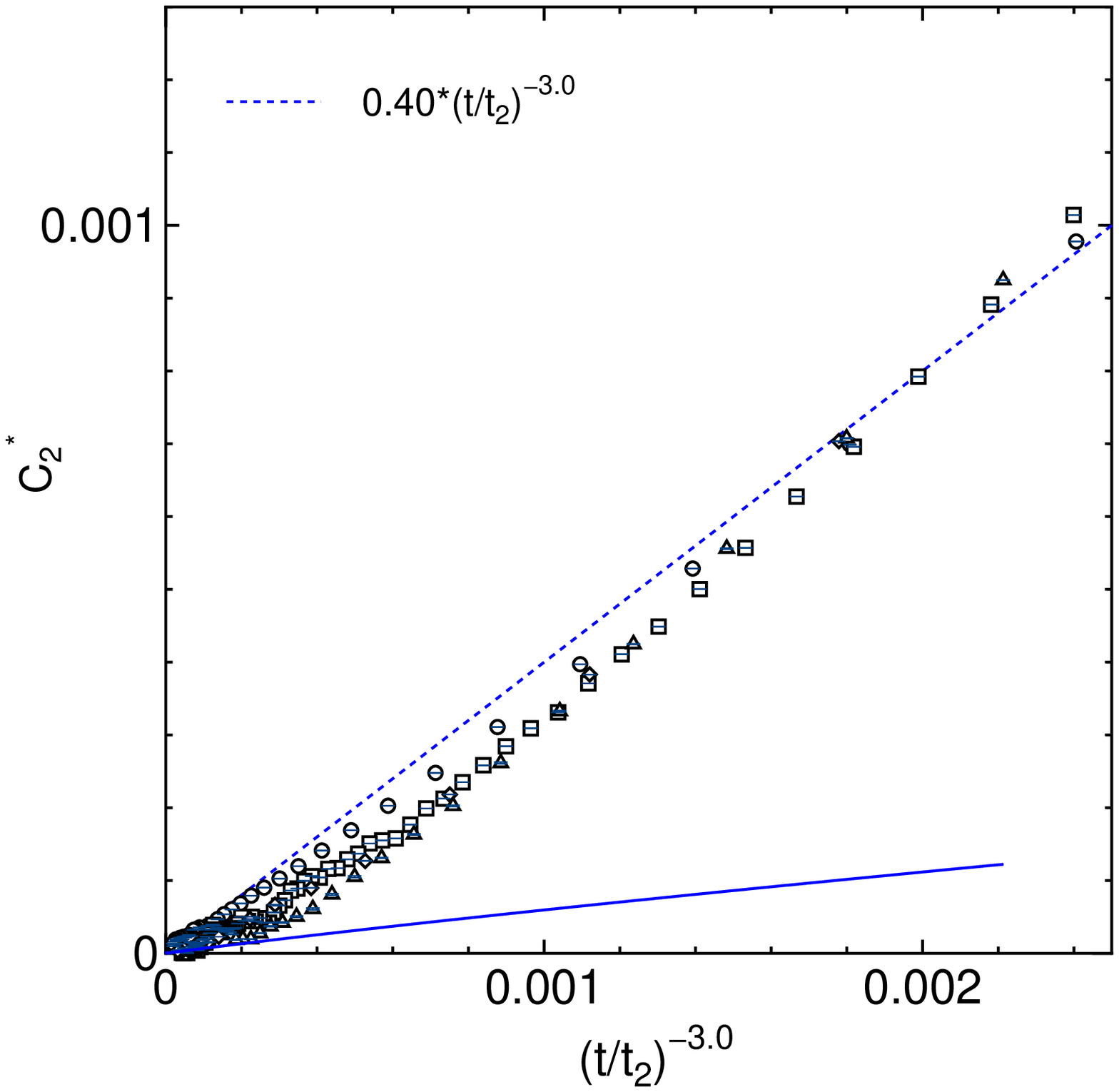}
\end{center}
\end{minipage}
\hspace{2mm}
\begin{minipage}{0.47\hsize}
\begin{center}
\includegraphics[scale=0.42]{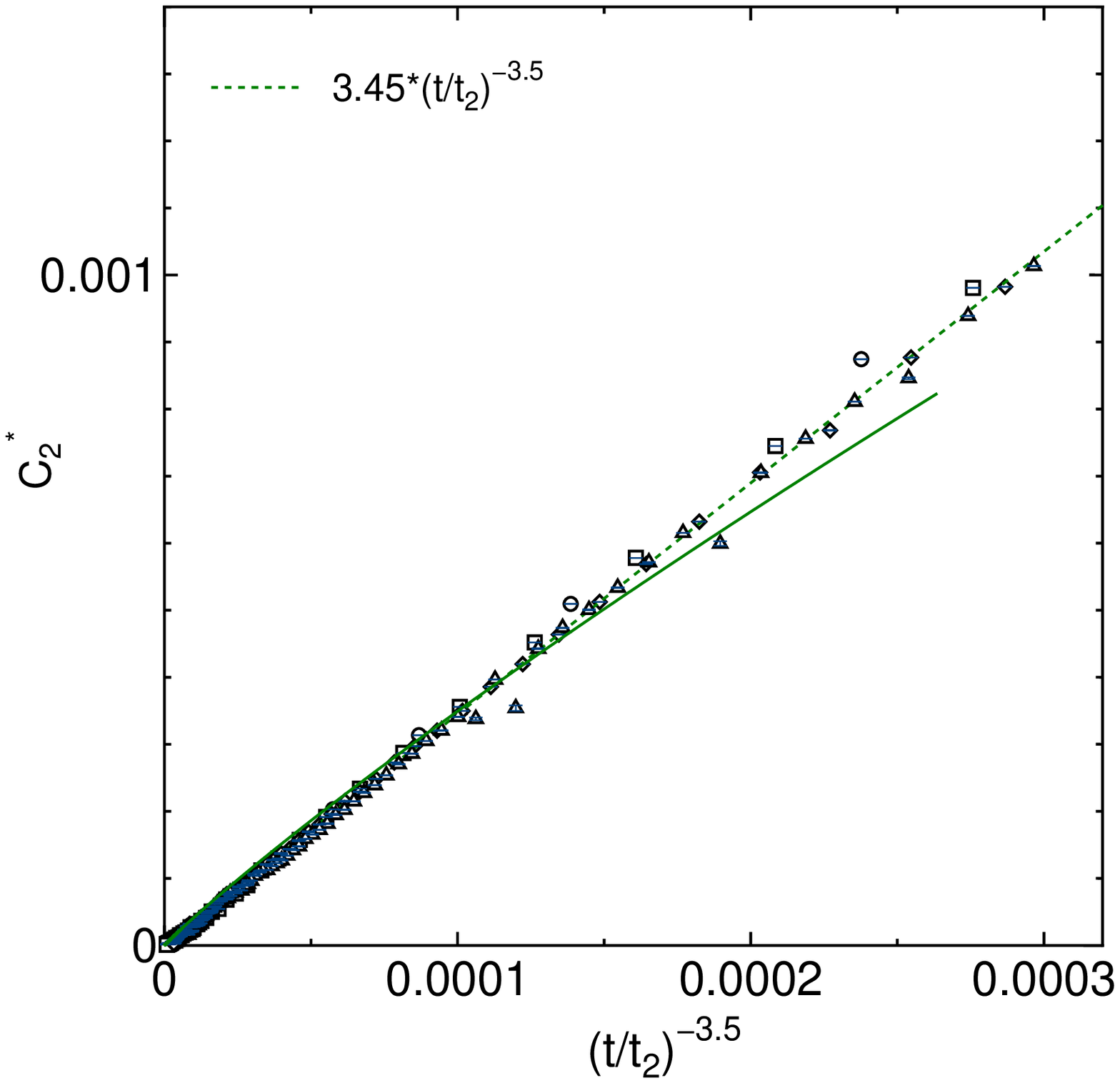}
\end{center}
\end{minipage}
\caption{Decay of $C^*_2$ in the diffusional power decay regime in terms of $(t/t_2)^{-\beta}$ for the several packing fractions in 2D(left) and 3D(right) for $4096$ and $512$ particles are shown.
The solid lines are theoretical results for $\nu=0.57$ in 2D(left) and $\nu=0.47$ in 3D(right), respectively. The dot lines are the linear fitting lines.}
\end{figure}

\noindent
Plotting $C^*_2$ in terms of $(t/t_2)^{-7/2}$ in Fig.~7, in the 3D case the data lead to a linear function, which indicates consistency only with the exponent of theoretical prediction of $\beta=7/2$ in the long time limit.
In the 2D case, numerical data suggest the exponent deviates from $\beta=3$.
Because of the divergence of the diffusion coefficient in 2D, the entire theory needs to be reworked.
The amplitudes of the simulation data in 3D are fairly good agreement with the theoretical results incorporating recollision contributions and a potential of mean force through the extra factor of $g(\sigma)$, (i.e., $(25/9)g(\sigma)C_2^*(t)$ in the long time limit), as shown in Fig.~3, 6 and 7, which are described in Ref.~\cite{leegwater_1989}.
In Table II, the values of amplitude $B'$ of eq.~(3.2) are shown.
We also show $B'$ in terms of scaled packing fraction ($\nu/\nu_c$: $\nu_c$ is solid-fluid transition point) in Fig.~6.

\begin{table}
\begin{center}
\begin{tabular}{cccccc} \hline \hline
packing fraction(2D) & $\nu$  & 0.57 & 0.65 & 0.67 & 0.69 \\
Amplitude  & $B'$ & 1.42 & 17.8 & 75.5 & 1849 \\ \hline
packing fraction(3D) & $\nu$  & 0.35 & 0.40 & 0.45 & 0.47 \\
Amplitude & $B'$ & 0.12 & 0.47 & 3.16 & 8.57 \\
Amplitude(Theory) & $B'$ & 0.047 & 0.32 & 4.00 & 11.58 \\ \hline \hline
\end{tabular}
\end{center}
\caption{
The amplitude $B'$ for various packing fractions are given and plotted in Fig.~6.}
\label{table:1}
\end{table}

\section{Discussion}

The cause of the stretched exponential relaxation in the molasses regime is considered to be due to the distribution of different life times of transient clusters of nuclei in dense fluid systems.
Thus, there exist several exponential relaxation decays for each collision pair.
To confirm the above speculation for the origin of the molasses effect in 2D at the microscopic level, we calculated the bond orientational order, $\phi_6$ as an alternative to $C_2$, and furthermore, to visualize the distribution of crystal clusters in the hard disk system.
The usual bond orientational order parameter $\phi_6$ for each hard disk $i$ is defined by

\begin{equation}
\phi_6^i = \left|\frac{1}{N_i} \sum_{j=1}^{N_i} \exp{(6i\theta_{ij})}\right|,
\end{equation}

\noindent
where $N_i$ is the number of the nearest neighbors around the tagged particles $i$, and $\theta_{ij}$ is the angle between the position vector from the disk $j$ to $i$ and an arbitrary fixed reference axis (e.g., $x$-axis).
Two disks are defined as neighbors if the distance is less than $\sqrt{3} \sigma$.
In this definition, $\phi_6^i$ takes on values between $0$ and $1$, which measures the degree of crystallization in terms of considering only nearest neighbors.

\begin{figure}
\begin{minipage}{0.50\hsize}
\begin{center}
\includegraphics[scale=0.52]{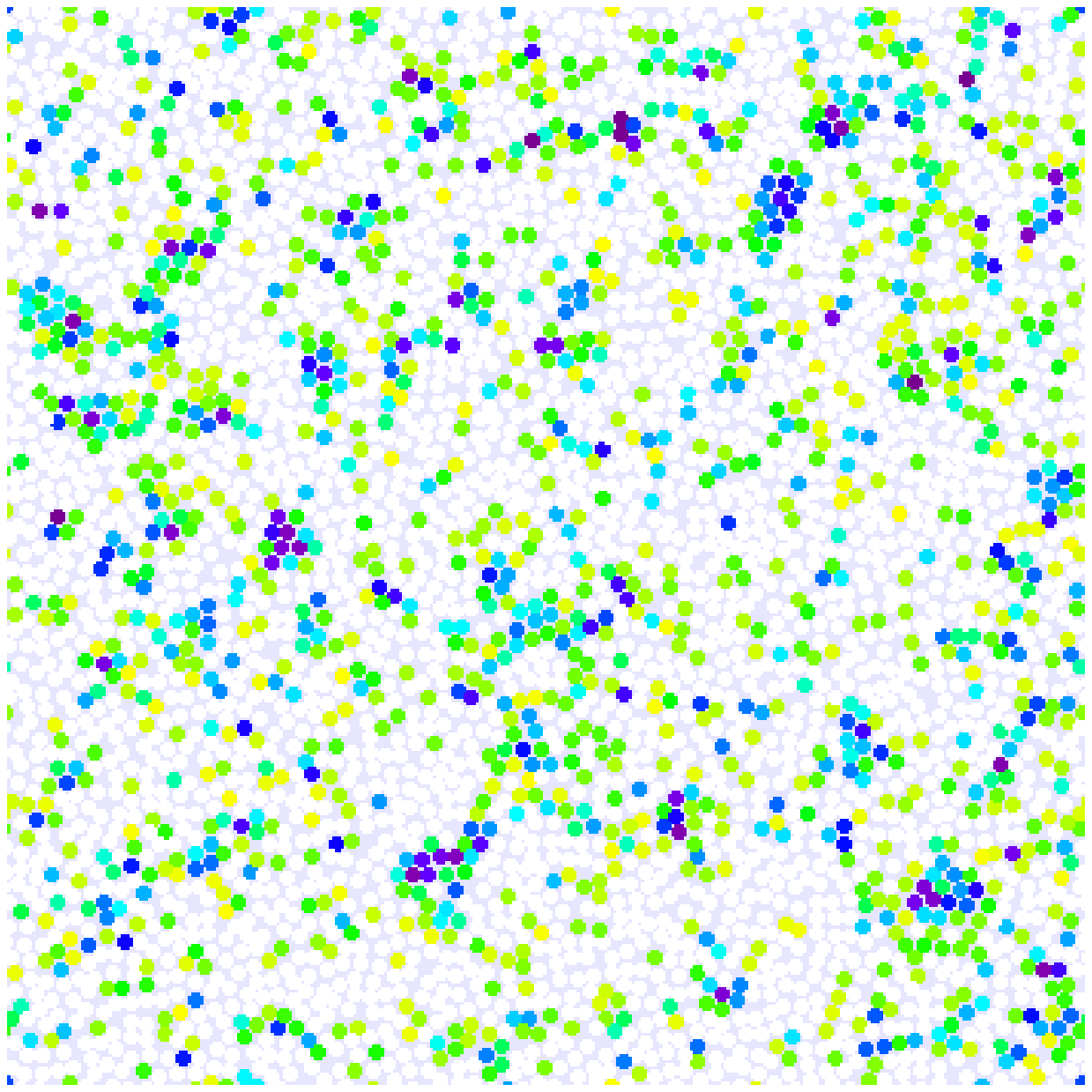}
\end{center}
\end{minipage}
\begin{minipage}{0.50\hsize}
\begin{center}
\includegraphics[scale=0.54]{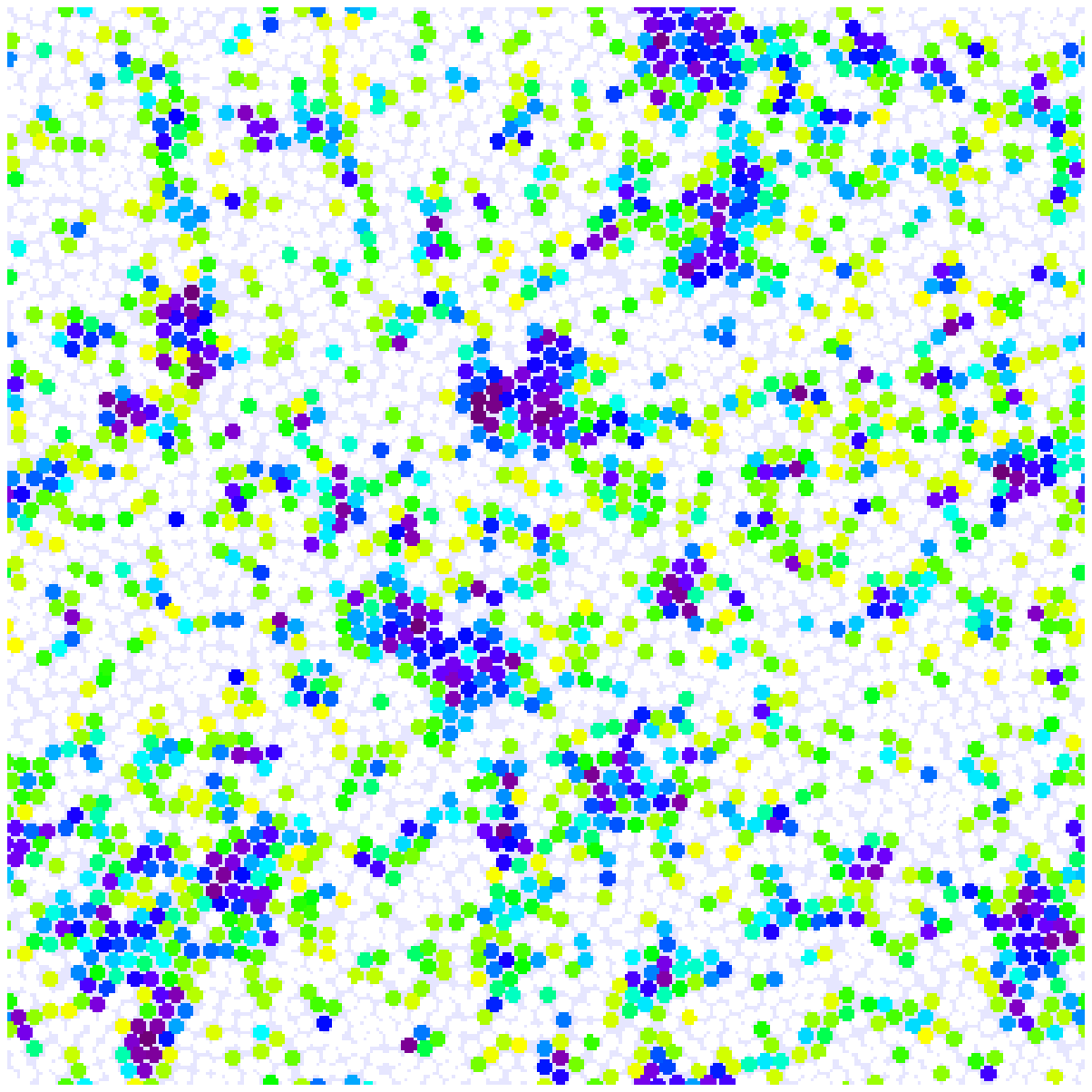}
\end{center}
\end{minipage}
\caption{The spatial distribution of $\phi_6^i$ for $4096$ particles system at a given time for two packing fractions, $\nu=0.65$(left) and $0.69$(right).
The darker the region, the closer $\phi_6$ is to unity.}
\end{figure} 

Figure 8 shows typical snapshots of the spatial distribution of $\phi_6^i$.
The gradation in shading of the particles indicate the value of $\phi_6^i$, the darker, the closes to unity.
We clearly observe the dramatic growth of several solid nuclei as the density nears solidification.
The averaged values of $\overline{\phi_6^i}$ and the number of particle with $0.9 < \phi_6^i < 1$ for each packing fraction are summarized in Table III.

\begin{table}
\begin{center}
\begin{tabular}{cccccc} \hline \hline
packing fraction(2D) & $\nu$  & 0.57 & 0.65 & 0.67 & 0.69 \\
averaged bond orientational order & $\overline{\phi_6^i}$ & 0.36 & 0.42 & 0.44 & 0.47 \\ 
particle number with $0.9 < \phi_6^i < 1$ & $N_s$ & 11 & 58 & 82 & 131 \\ 
\hline \hline
\end{tabular}
\end{center}
\caption{
The averaged bond orientational order $\overline{\phi_6^i}$ and the number of particle with $0.9< \phi_6^i <1$ for various packing fractions are shown.}
\label{table:1}
\end{table}

Understanding the slow decaying process of the OACF is considered a key factor in understanding the onset of the glass transition, and is here qualitatively confirmed to be due to the transitory existence of solid nuclei in 2D in Fig.~8.
Recently, their existence was also demonstrated in colloid experiments~\cite{conrad_2006}.
For quantitative purposes it is crucial to determine the spatial extents of the crystal nuclei and how long they persist.
In that direction we quantitatively determined in this paper that the decay of the pair distribution of $C_2$ could be characterized by a relaxation time $t_2$ at intermediate time (molasses regime) and later on by a pair diffusional breaking up of the clusters.
At the quantitative crossover time shown explicitly in Table I and Fig.6, the largest cluster at the freezing density of only a few sphere diameter in size persist for $\sim 431 t_0$, which corresponds for typical argon parameters estimate of $t_0 \sim 6.4 \times 10^{-14}$ [s] to only about 30 picoseconds ($\sim 2.8 \times 10^{-11}$[s]) in real liquids.
To make the further quantitative progress, we need to investigate the orientation correlation function of the quadruplet component, $C_4$, as a function of the distance between the two colliding pairs.
When this is carried out as a function of density, it could be possible to tell how the cluster size distribution changes and how long they last as the solidification density is approached, and, subsequently, determine how fast one has to increase the density to get a glass instead of a crystal.
This is however, computationally a very demanding task.
Instead we are planing to investigate an extension of $\phi_6^i$ to a higher order orientational parameter involving further neighbors and thus, hopefully, improve the methodology of an alternative efficient calculation for $C_4(t)$.

\section*{Acknowledgements}

We would like to thank Professor A.J.C. Ladd for helpful comments.
We also wish to thank Professors H. Mori, H. van Beijeren, W.G. Hoover, D. Frenkel, N. Ito and J. Wakou for valuable discussions.
This study was supported by Grant-in-Aid for Scientific Research from the Ministry of Education, Culture, Sports, Science and Technology No. 19740236.
Part of the computations for this study was performed using the facilities of the Supercomputer Center, Institute for Solid State Physics, the University of Tokyo, and Research Center for Computational Science, Okazaki, Japan.

\end{document}